\documentclass[aps,prl,twocolumn,superscriptaddress]{revtex4}
\usepackage{epsf,graphicx}
\usepackage{amssymb}
\usepackage{amsmath}
\usepackage{latexsym,bm,array,amsfonts,multirow}
\usepackage{color}
\usepackage{ulem}
\usepackage{epstopdf}
\begin{document}
	
	\title{Emergent superconducting stripes in two-orbital superconductors}
	\author{Qiong Qin}
	\affiliation{New Cornerstone Science Laboratory, Department of Physics, School of Science, Westlake University, Hangzhou 310024, Zhejiang, China}
	\author{Yi-feng Yang}
	\email[]{yifeng@iphy.ac.cn}
	\affiliation{Beijing National Laboratory for Condensed Matter Physics and Institute of
		Physics, Chinese Academy of Sciences, Beijing 100190, China}
	\affiliation{University of Chinese Academy of Sciences, Beijing 100049, China}
	\affiliation{Songshan Lake Materials Laboratory, Dongguan, Guangdong 523808, China}
	\date{\today}
	
	\begin{abstract}
Motivated by recent experiments in KTaO$_3$/EuO interface, we propose an intrinsic mechanism where superconducting stripes emerge naturally without involving disorder, charge inhomogeneity, or competing orders. Our theory is based on a two-orbital model of superconductivity, where one orbital displays a quasi-one-dimensional dispersion and the other orbital is more localized and contributes pairing interactions along the perpendicular direction. Our auxiliary-field Monte Carlo simulations demonstrate that the pairing amplitude exhibits spatial modulation such that the superconductivity naturally disaggregates into two-leg or three-leg superconducting stripes separated by non-superconducting blocks. Our work provides a promising scenario of emergent superconducting stripes in homogeneous two-dimensional systems and reveals unexpectedly rich physics in two-orbital superconductors for future materials design.
	\end{abstract}
	\maketitle

\textit{Introduction.--} 
Superconducting stripes have been extensively studied in correlated superconductors, e.g., the high-$T_c$ cuprates, and recently reported in SrTa$_2$S$_5$ and KTaO$_3$/EuO interface \cite{Tranquada2008,Devarakonda2024a,Hua2024}. While such phenomena have often been attributed to extrinsic factors such as disorder, strain, or competing magnetic/charge orders, the underlying mechanisms remain inconclusive.  In doped Mott insulators, for example, kinetic energy frustration may drive hole-rich rivers bordered by spin-ordered domain. In KTaO$_3$/EuO interface, however, no evidence of disorder or charge inhomogeneity has been reported in experiment. An intrinsic mechanism is therefore demanded under which superconductivity may naturally disaggregate into stripes.

In this work, we propose a scenario where superconducting stripes emerge without involving disorder, charge inhomogeneity, or competing orders. Our theory was initially motivated by recent studies of high-$T_c$ bilayer nickelate superconductors \cite{Sun2023b,Hou2023,Zhang2023c,Wang2024b,Li2024,Zhu2024,Zhang2023g}. While their detailed pairing mechanism is still under debate \cite{Shen2023Effective,Yang2023Possible,Lechermann2023Electronic,Gu2023,Qu2024,Lu2024,Heier2024,Fan2024,Chen2024b,Christiansson2023,Jiang2024,Sakakibara2024,Xue2024,Tian2024a,Sakakibara2024a,Jiang2024a,Wanghighly2025,Qin2023High,Qin2024a,Yang2023Inter,WangFermi2025,Yang2025a}, the low-lying physics is generally recognized to arise from two orthogonal orbitals: the almost half-filled $d_{z^2}$ orbital and nearly quarter-filled $d_{x^2-y^2}$ orbital \cite{Cao2024,Luo2023,Li2017,Chen2024a,Yang2024a,Liu2024c,Dong2024,Chen2024,Tian2024}, where the $d_{z^2}$ orbitals contribute inter-layer pairing interaction and the $d_{x^2-y^2}$ orbitals show good metallic behavior and hybridize with the $d_{z^2}$ orbitals \cite{Qin2023High,Qin2024a,Yang2023Inter}. As a result, Cooper pairs form along the $z$ axis but hop within the $xy$ plane to attain the phase coherence, in sharp contrast to most other superconductors. A phenomenological Ginzburg-Landau analysis has shown that the superconductivity in ideal multilayer ($\ge 4$) systems may naturally decompose into bilayer or trilayer superconducting blocks and exhibit order parameter modulation along the $z$ axis \cite{Yang2024}. However, valence difference or layer imbalance might prevent its occurrence in multilayer nickelates.

In KTaO$_3$/EuO interface \cite{Hua2024}, first-principles calculations also reveal two orthogonal orbitals around the Fermi energy: the spin-polarized $d_{xy}$ orbital and the itinerant $d_{yz}$ orbital with quasi-one-dimensional dispersion. These led us to propose an effective two-orbital model in two dimension, where the more localized orbital contributes the pairing along one direction and the more itinerant orbital provides the hopping along the other direction. The two orbitals hybridize between nearest-neighbor sites, which may help induce phase coherence for the superconductivity. In real materials, the model may be realized in multi-chain or square-lattice forms.

Our Monte Carlo simulations confirm the emergence of superconducting stripes in such two-orbital superconductors. Notably, even-chain and square-lattice systems predominantly consist of two-leg superconducting stripes separated by non-superconducting blocks, while odd-chain systems may undergo a phase transition from an entirely two-leg-stripe configuration to one with an additional three-leg stripe. In all cases, superconducting stripes emerge and remain robust without involving disorders, charge inhomogeneity, or competing orders, thus providing a potential explanation of the superconducting stripes observed in two dimensional interface. Our work opens a direction of two-orbital superconductivity for future materials design and experimental investigations.

\textit{Model and Method.--} We consider the two-orbital effective $t$-$V$-$J$ Hamiltonian in multi-chain form \cite{Yang2023Inter,Qin2023High,Wanghighly2025}:  
\begin{eqnarray}
	H&=&-t\sum_{lis}(c_{lis}^{\dagger}c_{l,i+1,s}+h.c.)-\mu\sum_{is}c_{lis}^{\dagger}c_{lis}\nonumber\\
	&&-V\sum_{lis}\left(c_{lis}^{\dagger}d_{l,i+1,s}+d_{lis}^{\dagger}c_{l,i+1,s}+h.c\right) \label{eq:H} \\
	&&-t_{\perp} \sum_{ais} \left(d_{ais}^\dagger d_{a+1,is}+h.c.\right)+J\sum_{ai}\bm{S}_{ai}\cdot\bm{S}_{a+1,i}, \nonumber
\end{eqnarray}
where $d_{lis}$ ($c_{lis}$) denotes the more localized (itinerant) orbital of spin $s$ on site $i$ of chain $l=1,~2,~\cdots~ N$, $\bm{S}_{li}=\frac12\sum_{ss'}d_{lis}^{\dagger}\bm{\sigma}_{ss'}d_{lis'}$ is the spin operator of the $d$ electrons, where $\bm{\sigma}$ are the Pauli matrices, $J$ ($t_\perp$) is the superexchange  interaction (hopping) of $d$ electrons between nearest-neighbor chains with $a=1,~2,~\cdots,~N-1$, $t$ is the nearest-neighbor hopping of $c$ electrons along the chain, $\mu$ is their chemical potential, and $V$ is the hybridization between two orbitals on nearest-neighbor sites in each chain. All parameters should be considered to have been renormalized by Coulomb repulsions. For simplicity, we assume spin-singlet pairing and ignore all other instabilities. This allows us to utilize the static auxiliary-field Monte Carlo approach, which has been applied to many correlated systems \cite{Mayr2005a,Dubi2007,Karmakar2020,Pasrija2016,Dong2021a,Mukherjee2014,Dong2022PRB,Qin2023PRB,Zhong2011,Singh2021,Qin2025npjQM} and predicted the correct magnitude of $T_c$ in bilayer and trilayer nickelate superconductors \cite{Qin2023High,Qin2024a}. 

\begin{figure}[tb]
	\begin{center}
		\includegraphics[width=8cm]{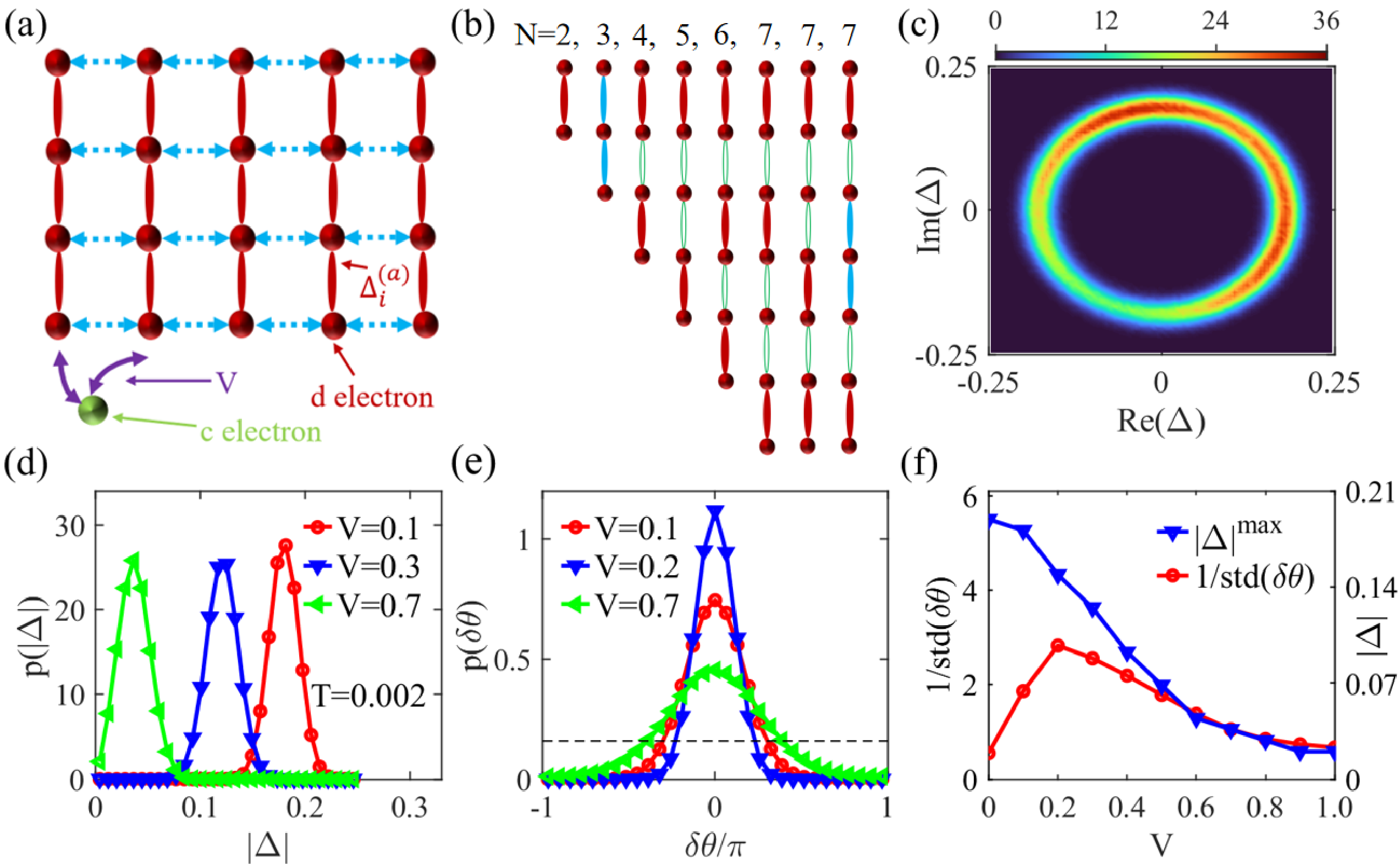}
	\end{center}
	\caption{	(a) Schematic representation of the two-orbital model.  
		(b) Illustrations of the stripe configurations for multi-chain systems of $2\le N\le7$. 
		(c) Intensity plot of the probabilistic distribution $p(\Delta)$ of the local pairing fields on complex plane at $V = 0.1$ for $N=2$.  
		(d) Probabilistic distribution $p(|\Delta|)$ of the pairing field amplitude for different $V$.  
		(e) Corresponding phase difference distribution $p(\delta\theta)$ of the pairing fields between two distant bonds for different $V$.  
		(f) Evolution of the peak position $|\Delta|^{\rm max}$ in $p(|\Delta|)$ and the inverse standard deviation, $1/{\rm std}(\delta \theta)$, derived from $p(\delta\theta)$ with the hybridization $V$. All data were obtained at $T = 0.002$.
	}
	\label{fig1}
\end{figure}

To perform numerical simulations, we first decouple the superexchange term \cite{Coleman2015}:  
\begin{equation}
	J\bm{S}_{ai}\cdot\bm{S}_{a+1,i}\rightarrow \sqrt{2}\bar{\Delta}_i^{(a)}\Phi_{i}^a+\sqrt{2}\bar{\Phi}^a_{i}\Delta_i^{(a)}+\frac{8\bar{\Delta}_i^{(a)}\Delta_i^{(a)}}{3J},
\end{equation}
where $\Phi_{i}^a=\frac{1}{\sqrt{2}}\left(d_{ai\downarrow}d_{a+1,i\uparrow}-d_{ai\uparrow}d_{a+1,i\downarrow}\right)$ represents the inter-chain spin singlet of $d$ electrons at site $i$ between $a$-th and $(a+1)$-th chains and $\Delta_i^{(a)}$ is the corresponding auxiliary pairing field, as illustrated in Fig.~\ref{fig1}(a). To avoid the sign problem, we neglect the imaginary time dependence of the auxiliary fields, $\Delta_i^{(a)}(\tau)\rightarrow \Delta_i^{(a)}=|\Delta_i|^{(a)}e^{i\theta_i^a}$, which retains full thermodynamic and spatial fluctuations, thus describing well superconducting phase fluctuations beyond mean-field approximation \cite{Qin2025npjQM}. The resulting bilinear form allows us to integrate out all fermionic degrees of freedom and obtain an effective action, $S_{\rm eff}(\{\Delta_i^{(a)}\})$, solely of the pairing fields. Monte Carlo simulations are then performed to sample the probabilistic distribution $p(\{\Delta_i^{(a)}\})=Z^{-1}e^{-S_{\rm eff}(\{\Delta_i^{(a)}\})}$, where $Z$ is the partition function serving as the normalization factor. For simplicity, we set $t=1$ as the energy unit, fix $\mu=-1.3$, $J=0.5$, $T=0.002$, and tune $V$ as the free parameter. For multi-chain systems, we apply periodic boundary condition along the chain direction and open boundary condition  perpendicular to the chains. For square lattice, periodic boundary conditions are applied to both directions. The total number of lattice sites ranges from 196 to 210 depending on the lattice geometry. All results have been verified to be qualitatively consistent for larger lattice sizes.

\begin{figure}[t]
	\begin{center}
		\includegraphics[width=8cm]{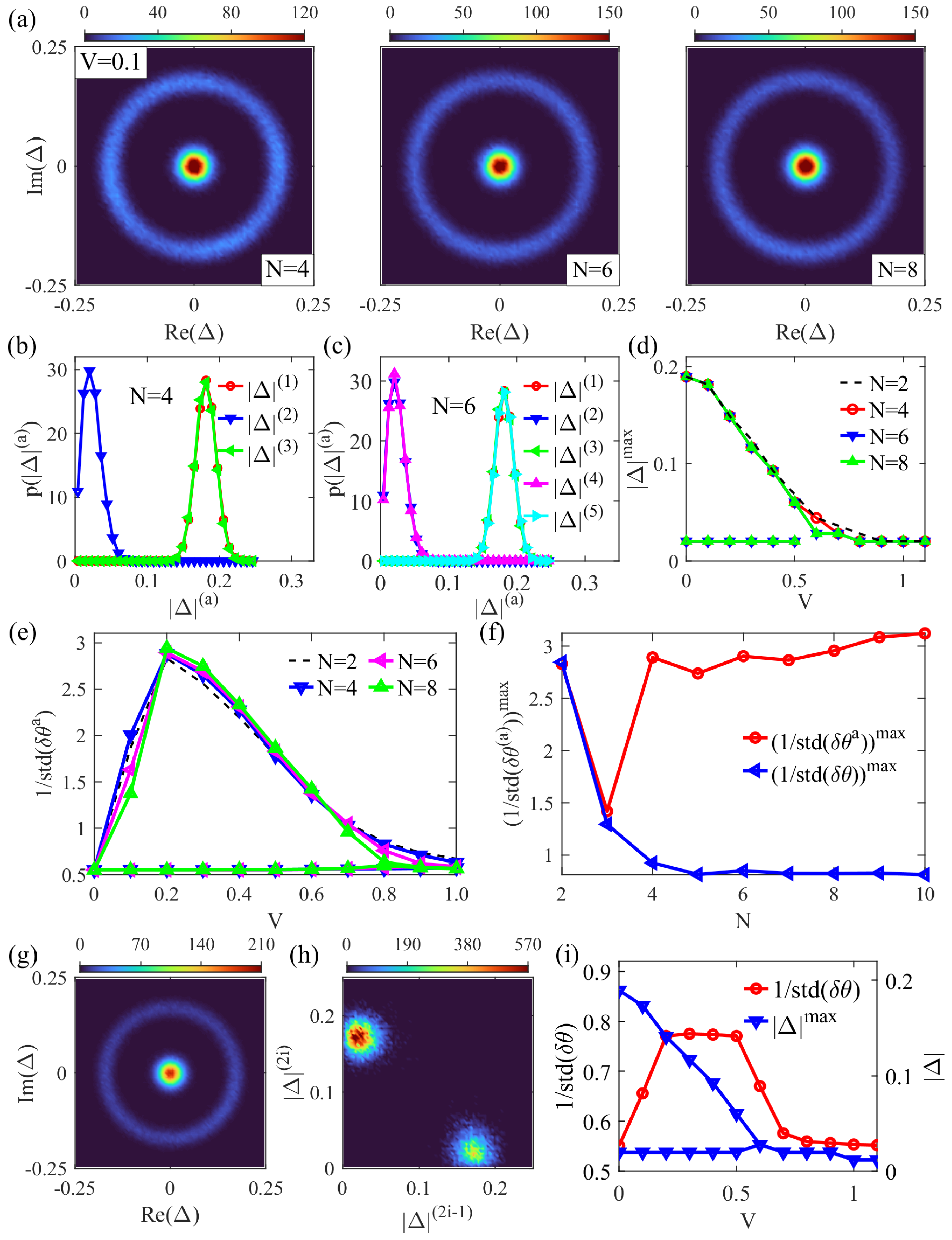}
	\end{center}
	\caption{(a) Intensity plots of the probabilistic distribution $p(\Delta)$ of the local pairing fields at $V=0.1$ for $N = 4$, $6$, $8$.  
	(b) and (c) are the corresponding amplitude distribution $p(|\Delta|^{(a)})$ for $N = 4$ and $6$, respectively.
	(d) Peak position $|\Delta|^{\rm max}$ of $p(|\Delta|)$ as a function of the hybridization $V$ for $N = 4$, $6$, $8$.  
	(e) Evolution of the inverse standard deviation $1/{\rm std}(\delta \theta^a)$ of the phase difference between distant bonds in superconducting and non-superconducting blocks separately. 
	 (f) Comparison of the maximum $1/{\rm std}(\delta \theta^{(a)})$ of superconducting blocks and those averaged over all blocks ($1/{\rm std}(\delta \theta)$) as functions of $N$.
	 (g) Intensity plot of $p(\Delta)$ at $V = 0.1$ for the square lattice model.
	 (h) Joint distribution of the pairing amplitudes between two adjacent blocks.
	 (i) $|\Delta|^{\rm max}$ and $1/{\rm std}(\delta \theta)$ (calculated over all blocks) as functions of the hybridization $V$. The data in (g-i) were produced on a $14\times14$ square lattice.
}
	\label{fig2}
\end{figure}

\textit{$N=2$.--} As a reference for larger $N$, we first study the case $N=2$, a two-leg ladder with pairing interaction on the rung and hopping along both legs. Figure \ref{fig1}(c) plots the distribution $p(\Delta)$ on the complex plane $(\rm{Re}\Delta,\rm{Im}\Delta)$ for all $\Delta_i$ at $V=0.1$. The ring shape indicates superconducting phase fluctuations without breaking the global U(1) symmetry at finite temperature. The amplitude distribution $p(|\Delta|)$ is given in Fig.~\ref{fig1}(d), showing peaks at $|\Delta|^{\rm max}$ moving towards smaller values as $V$ increases. Figure \ref{fig1}(e) plots the distribution of the phase difference $\delta \theta$ between two distant bonds separated by an extended intra-chain distance of 10 sites. The peak shape centered around $\delta\theta=0$ and its width reflect phase correlation and fluctuations along the chain, whose standard deviation, ${\rm std}(\delta\theta)\equiv \sqrt{\langle(\delta\theta)^2\rangle}$, is inversely related to the superfluid density. As $V$ increases, the peak gets narrowed but then broadened, causing nonmonotonic evolution of $1/{\rm std}(\delta\theta)$ plotted in Fig.~\ref{fig1}(f), in contrast to the continuous suppression of $|\Delta|^{\rm max}$. This implies a dual role of the hybridization, which tends to induce the phase coherence but at the same time reduce the pairing strength. As a result, the superconductivity at large $V$ is constrained by the pairing strength, as expected in the weak-coupling BCS theory, but at small $V$ is governed by phase fluctuations, as studied extensively in underdoped cuprate superconductors \cite{Keimer2015a,Emery1995a}.

\begin{figure}[t]
	\begin{center}
		\includegraphics[width=8cm]{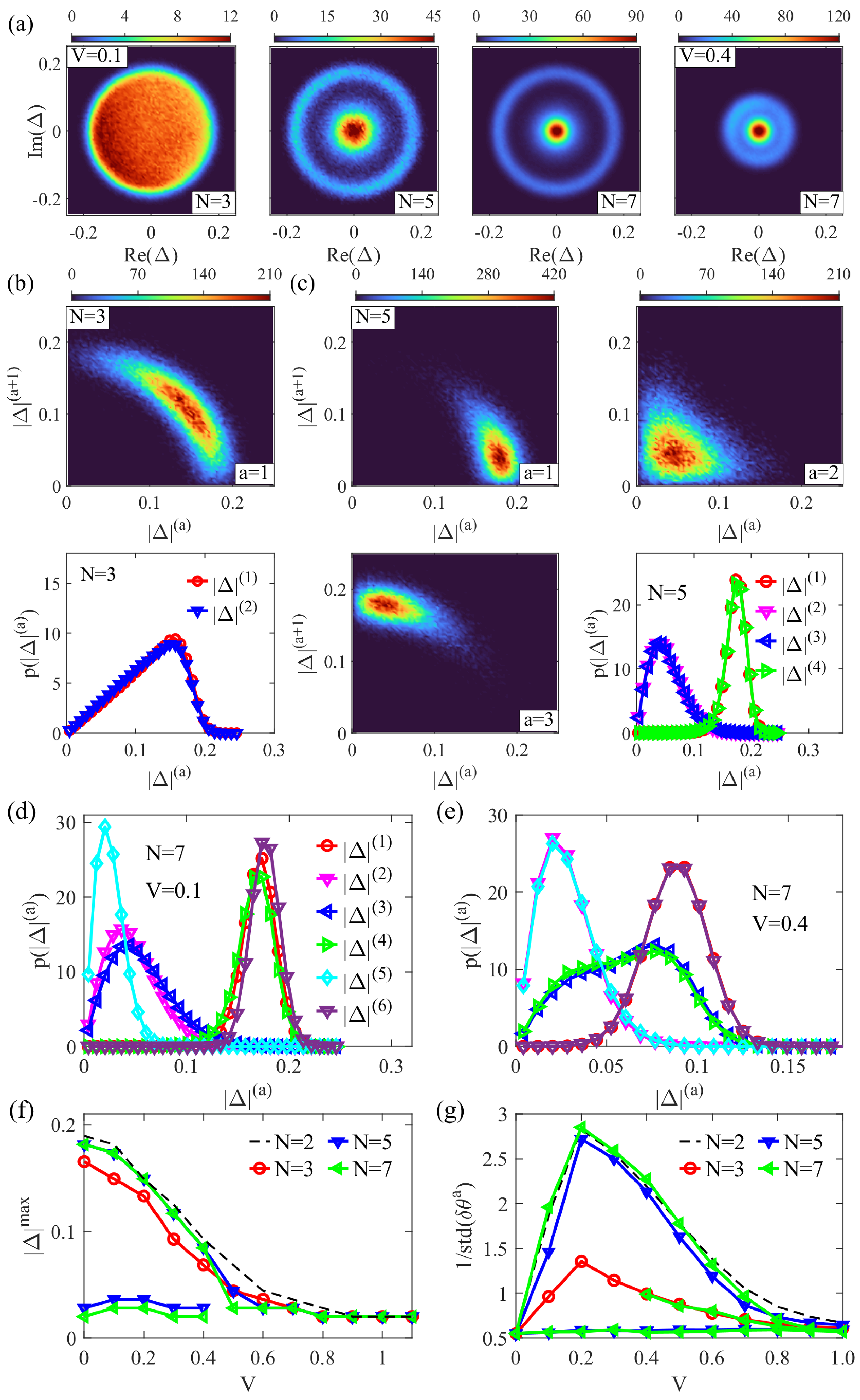}
	\end{center}
	\caption{(a) Intensity plot of $p(\Delta)$ for $N = 3$, $5$, $7$.  
	(b) and (c) show the amplitude distribution $p(|\Delta|^{(a)})$ of the local pairing fields and their joint distribution $p(|\Delta|^{(a)}, |\Delta|^{(a+1)})$ between two nearest-neighbor blocks for $N = 3$ and $5$, respectively.  
	(d) and (e) present $p(|\Delta|^{(a)})$ for $N = 7$ at $V = 0.1$ and $0.4$, respectively. 
	(f) and (g) compare the extracted $|\Delta|^{\rm max}$ and ${\rm std}(\delta \theta^{(a)})$ with varying hybridization $V$ for $N = 2$, $3$, $5$, $7$.
}
	\label{fig3}
\end{figure}

\textit{Even $N$.--} Figure \ref{fig2}(a) shows $p(\Delta)$ on complex plane for $N = 4$, 6, 8 at $V = 0.1$. Inside the ring, an additional disk shape appears at the origin, implying thermal fluctuations without pairing. We find two well-separated peaks in $p(|\Delta|^{(a)})$ for $N = 4$, $6$ in Figs. \ref{fig2}(b) and \ref{fig2}(c), respectively, where $|\Delta|^{(1, 3, 5)}$ correspond to the ring and $|\Delta|^{(2, 4)}$ the disk. This indicates spatial modulation of the pairing amplitudes with two-leg superconducting stripes separated by non-superconducting blocks, as illustrated in Fig. \ref{fig1}(b) for $N=4$, 6. This pattern persists for all even-chain systems and extends to the square lattice as shown in Figs. \ref{fig2}(g) and \ref{fig2}(h), where it breaks the lattice translational symmetry along the pairing direction. Figure \ref{fig2}(d) shows $|\Delta|^{\rm max}$ as a function of $V$. As $V$ increases, $|\Delta|^{\rm max}$, hence the pairing strength, of the superconducting stripes is reduced gradually, and the rest blocks remain non-superconducting. Again, as shown in Fig. \ref{fig2}(e), the inverse phase fluctuation $1/{\rm std}(\delta \theta^{a})$ of superconducting blocks varies nonmonotonically with $V$. Similar behaviors also exist in the square lattice model as shown in Fig. \ref{fig2}(i). Interestingly, as plotted in Fig. \ref{fig2}(f), the peak values, $(1/{\rm std}(\delta \theta^a))^{\rm max}$, of the superconducting blocks vary only slightly with $N$ except for $N=3$. This suggests that all two-leg superconducting blocks exhibit similar properties insensitive to the whole lattice. By contrast, those of the overall phase distributions decrease with increasing $N$ and saturate for $N\ge5$ due to the contribution from non-superconducting blocks, indicating stronger inter-block phase fluctuations.

\textit{Odd $N$.--} The odd $N$ cases display very different patterns. Figure \ref{fig3}(a) plots the distribution $p(\Delta)$ for $N = 3$, 5, 7. A broader disk is seen for $N = 3$. Correspondingly in Fig. \ref{fig3}(b), the pairing amplitudes of $a=1$ and 2 blocks exhibit almost identical distributions, with their joint distribution $p(|\Delta|^{(1)},|\Delta|^{(2)})$ forming a quarter-ring shape. This implies that both blocks are superconducting but fluctuate strongly around an overall amplitude, since both outer chains tend to form spin singlets with the inner chain. As a result, the phase difference fluctuations are also stronger, explaining the anomalous suppression of $(1/{\rm std}(\delta \theta^a))^{\rm max}$ for $N=3$ in Figs. \ref{fig2}(f) and \ref{fig3}(g). Similar competition has been proposed for trilayer nickelate superconductors, where the upper and lower layers compete to form spin-singlet pairs with the inner layer, causing a reduced $T_c$ compared to bilayer nickelates \cite{Qin2024a}.

The distribution $p(\Delta)$ for $N = 5$ appears similar to those of $N = 4$ and $6$, except that the central disk is now slightly broadened. Its amplitude distribution is shown in Fig. \ref{fig3}(c). We find that $|\Delta|^{(1,4)}$ are superconducting and $|\Delta|^{(2,3)}$ are non-superconducting, giving rise to two superconducting stripes on boundaries separated by a three-leg non-superconducting block in the middle, as illustrated in Fig. \ref{fig1}(b).

For $N=7$, the intensity plot of $p(\Delta)$ on complex plane reveals two distinct states at small and large $V$, with the transition occurring at around $V=0.35$ for $T=0.002$. Comparison with the amplitude distribution shown in Fig. \ref{fig3}(d) suggests that the state at small $V$ is similar to the $N=5$ case but with three two-leg superconducting stripes separated by a two-leg non-superconducting block and a three-leg non-superconducting block, while the state at large $V$ contains two two-leg superconducting stripes on  boundaries and one three-leg superconducting stripe (resembling the $N=3$ case) in the middle separated by two non-superconducting blocks, as illustrated in Fig. \ref{fig1}(b) for $N=7$. Similar analyses for larger odd $N$ confirm that two-leg superconducting stripes always emerge at both boundaries. Depending on the location of the three-leg non-superconducting block (small $V$) or three-leg superconducting stripe (large $V$) in the middle, the superconductivity contains $\frac{N - 3}{2}$ or $\frac{N - 5}{2}$ near degenerate configurations, respectively. In all cases, as shown in Figs. \ref{fig3}(f) and \ref{fig3}(g), the extracted $|\Delta|^{\rm max}$ exhibits similar suppression with increasing $V$, in contrast to the nonmonotonic variation of $1/{\rm std}(\delta \theta^{a})$, again reflecting the crossover between strong- and weak-coupling superconductivity. 

\begin{figure}[t]
	\begin{center}
		\includegraphics[width=8cm]{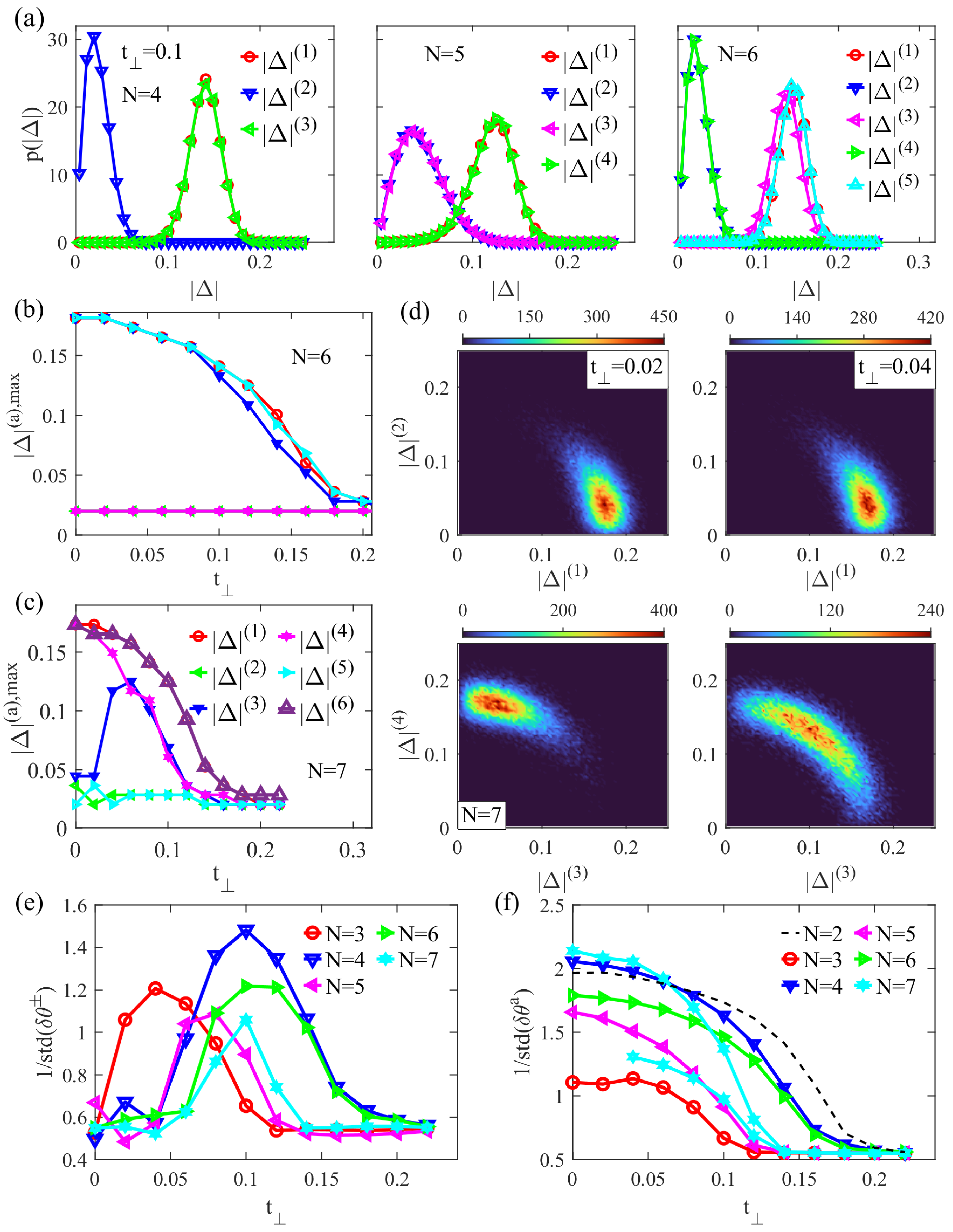}
	\end{center}
	\caption{(a) Amplitude distribution  $p(|\Delta|)$ of the local pairing fields for $N= 4$, $5$, $6$ at $t_{\perp}=0.1$. (b) and (c) show the corresponding peak position, $|\Delta|^{(a),\text{max}}$, of the block-dependent amplitude distribution varying with $t_\perp$ for $N = 6$ and $7$, respectively. (d) Joint distribution $p(|\Delta|^{(a)}, |\Delta|^{(a+1)})$ between two adjacent blocks at $t_{\perp}=0.02$, 0.04 for $N = 7$. (e) and (f) compare the inter-block $1/{\rm std}(\delta \theta^{\pm})$ and intra-block $1/{\rm std}(\delta \theta^a)$ of the superconducting stripes as functions of $t_{\perp}$ for different $N$. All results were obtained at $V=0.1$ and $T=0.002$.
}
	\label{fig4}
\end{figure}

\textit{Robustness of stripes.--} For simplicity, we have assumed $t_{\perp}=0$ for the correlated $d$ orbitals in above calculations. To examine the robustness of the emergent superconducting stripes, we discuss the effect of a finite $t_\perp$, whose value is supposed to be strongly renormalized by the onsite Coulomb repulsion and thus small. The results for $t_\perp=0.1$ are presented in Fig. \ref{fig4}, showing no significant difference in the distribution for $N\le 5$. For $N=6$, there appears a slight deviation between the peak positions $|\Delta|^{(a),\text{max}}$ of the inner and outer superconducting blocks, as is seen in Fig. \ref{fig4}(b) above $t_\perp=0.08$. For $N=7$, a finite $t_{\perp}$ tends to stabilize the three-leg superconducting stripe in the center and cause the transition between two superconducting states discussed earlier. This is clearly seen in Fig.~\ref{fig4}(c), where the peak position of $|\Delta|^{(3)}$ grows rapidly first and then merges with that of $|\Delta|^{(4)}$ as $t_\perp$ increases. Correspondingly in Fig.~\ref{fig4}(d), the joint distribution $p(|\Delta|^{(3)},|\Delta|^{(4)})$ for $N=7$ changes to a quarter-ring shape. This three-leg stripe turns non-superconducting at $t_\perp\approx 0.12$, while the two-leg superconducting stripes on two boundaries persist to a larger $t_\perp\approx0.18$, beyond which the superconductivity is completely destroyed. In all cases, non-superconducting blocks remain almost untouched, confirming the robustness of stripe formation.

The effect of $t_\perp$ may be further understood from Fig. \ref{fig4}(e), where $\delta \theta^{\pm}$ denotes the phase difference between two superconducting blocks. With increasing $t_{\perp}$, the inverse standard deviation $1/{\rm std}(\delta \theta^{\pm})$ first increases, marking the gradual development of phase coherence between different superconducting blocks, and then decreases following the continuous suppression of $1/{\rm std}(\delta \theta^{a})$ within each superconducting block shown in Fig. \ref{fig4}(e). Thus, while the inter-chain hopping promotes the inter-block phase coherence, it simultaneously reduces the coherence within each superconducting block. For large $t_\perp$, the latter constrains the superconducting phase coherence along the perpendicular direction, causing the nonmonotonic evolution of $1/{\rm std}(\delta \theta^{\pm})$ shown in Fig. \ref{fig4}(d). Only for sufficiently large $t_{\perp}$, the inter-block $1/{\rm std}(\delta \theta^{\pm})$ gets comparable to the intra-block $1/{\rm std}(\delta \theta^a)$, indicating similar phase coherence along both directions. At smaller $t_{\perp}$, the superconductivity remains anisotropic, as observed in experiments \cite{Tranquada2008,Devarakonda2024a,Hua2024,Xu2025,Cheng2025}.

\textit{Discussion and conclusions.--}We have proposed an intrinsic mechanism for emergent superconducting stripes without involving disorder, charge inhomogeneity, or competing orders. Our theory is based on a special two-orbital model with the pairing and hopping occurring along perpendicular directions. For even-chain systems or the square lattice, we show that the superconductivity naturally disaggregates into two-leg superconducting stripes separated by non-superconducting blocks, while for odd-chain systems, a phase transition occurs between different stripe patterns. In all cases, the stripe phenomena remain robust. Note that our conclusion does not depend on the special pairing type. It may therefore be readily extended to help understand the emergent superconducting stripes observed in homogeneous two-dimensional systems such as the KTaO$_3$/EuO interface \cite{Hua2024}. The unexpectedly rich physics revealed here will also stimulate future materials design and experimental investigations of two-orbital superconductors.

This work was supported by the National Natural Science Foundation of China (Grants No. 12474136 and No. 12447125).

\end{document}